\documentclass[twocolumn]{aastex701}
\usepackage{float}
\usepackage{caption} 
\usepackage{amsmath}
\usepackage{graphicx}
\begin{document}

\title{Candidate O I 7774 Å Absorption from an Escaping Rock-Vapor Outflow of BD+05 4868 Ab}

\correspondingauthor{Kody M. Walker}

\author[0009-0000-0948-1320,sname=Walker]{Kody M. Walker}
\affiliation{Department of Chemistry and Physics,
Indiana State University,
Terre Haute, IN 47809, USA}
\email[show]{kwalker82@leomail.etamu.edu}

\author[sname=Wallace,suffix=II]{James C. Wallace II}
\affiliation{Department of Earth and Environmental Systems,
Indiana State University,
Terre Haute, IN 47809, USA}
\email{James.Wallace@indstate.edu}
\begin{abstract}
{Disintegrating rocky planets provide a rare opportunity to probe
material escaping directly from intensely irradiated planetary
surfaces. We present time-resolved, low-resolution
($R\simeq600$) optical spectroscopy and complementary ground-based
photometry of BD+05~4868~Ab. The in-transit spectra exhibit an
additional unresolved absorption feature near the O~{\sc i} triplet
at 7771.94, 7774.17, and 7775.39~\AA\ relative to the out-of-transit
baseline. The feature has an equivalent width of
$W_\lambda\simeq1.4\pm0.2$~\AA, a maximum differential depth of
approximately 9\%, and an apparent width of 13--16~\AA, comparable
to the instrumental resolution. We interpret the feature as candidate
O~{\sc i} absorption from an oxygen-bearing mineral-vapor outflow.
The $3s\,{}^5S^\circ_2$ lower level of the triplet lies 9.146~eV
above the ground state and cannot be appreciably populated in LTE at
the reported planetary temperature, requiring non-LTE excitation. A
one-zone forward model in which photoionization, O$^+$ recombination,
radiative cascade, and radiative trapping populate the excited state
reproduces the measured equivalent width and instrumentally broadened
line profile. The model favors an atomic component with a projected
extent larger than the narrow condensed-dust layer and permits
gas-to-dust mass-loss ratios of order a few for representative
ionizing irradiation. Stellar O~{\sc i} line-profile variations
remain the principal alternative at the present spectral resolution.
We therefore identify escaping O~{\sc i} as the preferred physical
interpretation while retaining the designation candidate O~{\sc i}.
High-resolution, phase-resolved spectroscopy can directly test this
interpretation by resolving the individual triplet components.}
\end{abstract}


\section{Introduction}
\label{sec:introduction}

{Rocky planets on ultra-short-period orbits can be heated to
temperatures at which silicate material vaporizes directly from their
surfaces. For sufficiently low-mass planets, this mineral vapor can
escape, expand, and partially condense into dust, producing the highly
variable and asymmetric transit profiles characteristic of
disintegrating rocky planets
\citep{Rappaport2012,PerezBecker2013,vanLieshoutRappaport2018,
Booth2023}. Systems such as KIC~12557548~b
(Kepler-1520~b), KOI-2700~b, and K2-22~b demonstrate that the
observed occultation is often dominated by extended dusty material
rather than by a compact planetary disk
\citep{Rappaport2012,Rappaport2014,SanchisOjeda2015}.
Photometric variability, wavelength-dependent extinction, and tail
morphology have consequently provided important constraints on grain
sizes, dust dynamics, and mass loss in these systems
\citep{Brogi2012,Croll2014,Bochinski2015,vanLieshout2014,
vanLieshout2016,RiddenHarper2018}.}

{The escaping material is not expected to consist of dust alone.
Thermochemical calculations of volatile-poor silicate vapor predict
gas-phase species including Na, O$_2$, O, SiO, Mg, Fe, and related
molecules and ions, with the composition depending on temperature,
bulk composition, fractional vaporization, and condensation
\citep{SchaeferFegley2009,Miguel2011,SchaeferLoddersFegley2012,
Kite2016,Costa2017}. Hydrodynamic calculations likewise show that
irradiated mineral outflows can contain neutral atoms, ions, and free
electrons while grains form farther from the source
\citep{ItoIkoma2021,Booth2023,CamposEstrada2024}. Gas and dust may
therefore represent different phases of the same escaping
rock-vapor flow.}

{Oxygen is particularly important because it is a major constituent of
silicate material and is expected to remain abundant in rocky vapor.
A schematic pathway connecting oxygen-bearing mineral vapor to the
excited neutral oxygen relevant to optical spectroscopy is}

\begin{equation}
{\rm O_2}
\xrightarrow{h\nu}
2{\rm O}
\xrightarrow{h\nu}
{\rm O^+}+e^-
\xrightarrow[\rm cascade]{e^-}
{\rm O~{\sc i}}(3s\,{}^5S^\circ)+\cdots .
\label{eq:oxygen_pathway}
\end{equation}

{This sequence is not intended to represent a complete chemical
network. Rather, it illustrates how photodissociation,
photoionization, recombination, and radiative cascade can connect an
oxygen-bearing mineral vapor to excited neutral oxygen
\citep{ItoIkoma2021,MaurerMazzali2010,Jerkstrand2015,Nahar2022}.}

{A particularly useful optical diagnostic is the O~{\sc i} triplet at
7771.94, 7774.17, and 7775.39~\AA. These transitions originate from
the excited $3s\,{}^5S^\circ_2$ lower level, which lies approximately
9.146~eV above the ground state
\citep{NISTASD2024,Nahar2022}. This large excitation energy makes the
triplet fundamentally different from a resonance line originating in
the ground state. In LTE, the lower-level population contains the
Boltzmann factor}

\begin{equation}
\frac{n_{5S}}{n_0}
\propto
\frac{g_{5S}}{g_0}
\exp\left(-\frac{E_{5S}}{k_{\rm B}T}\right).
\label{eq:boltzmann_intro}
\end{equation}

{For the reported substellar temperature of
BD+05~4868~Ab, $T\simeq1820$~K \citep{Hon2025},
$k_{\rm B}T\simeq0.157$~eV, giving}

\[
\exp\left(-\frac{9.146}{0.157}\right)
\simeq 4.7\times10^{-26}.
\]

{Direct thermal population of the $3s\,{}^5S^\circ$ state is therefore
negligible at the characteristic planetary temperature. Detectable
O~{\sc i} 7774~\AA\ absorption requires a non-LTE population
mechanism, with photoionization followed by O$^+$ recombination and
radiative cascade providing a natural pathway into the excited
quintet system. NLTE effects are well established for the
O~{\sc i} 777~nm triplet in stellar line formation
\citep{Takeda2003,Amarsi2018,Bergemann2021}, and the triplet has also
been detected in high-resolution exoplanet transmission spectroscopy,
where NLTE treatment was required to reproduce the observed line
strengths \citep{Borsa2022}. Although those systems are physically
different from a disintegrating rocky planet, they establish the
atomic and observational relevance of the transition.}

{BD+05~4868~Ab is particularly favorable for investigating an atomic
component of a rocky-planet outflow. The planet orbits a bright K
dwarf with a period of 1.271869~d and exhibits strongly variable
occultations with depths of approximately 0.8--2.0\%
\citep{Hon2025}. Modeling of the system reveals both leading and
trailing dust tails, characteristic grain sizes of approximately
$1$--$10~\mu$m, and a dust mass-loss rate of order
$\dot{M}_{\rm dust}\sim2\times10^{12}~{\rm g\,s^{-1}}$.
The trailing component is substantially more extended than the
leading tail, demonstrating that a large column of freshly supplied
rocky material persists around the planet. Radial-velocity
observations constrain the planetary mass to
$M_p<6.2\,M_\oplus$ at $5\sigma$; the much smaller
$\sim0.02\,M_\oplus$ value used in dust-tail simulations is a
fiducial model assumption rather than a measured mass
\citep{Hon2025}.}

{Recent observations demonstrate that disintegrating planets can carry
compositional information in both their condensed and gaseous
components. JWST mid-infrared observations of K2-22~b constrain the
mineralogy of its escaping dust and reveal spectral structure
potentially associated with gas-phase material
\citep{Tusay2025}. Ground-based spectroscopy has also searched for
atomic species such as Na and Ca in disintegrating systems
\citep{Gaidos2019,RiddenHarper2019}. Optical atomic spectroscopy
therefore provides a complementary route to probing the gas produced
alongside the dust.}

{In this work, we present time-resolved, low-resolution optical
spectroscopy and complementary ground-based photometry of
BD+05~4868~Ab. We identify a transit-correlated absorption feature
near the unresolved O~{\sc i} 7774~\AA\ triplet and adopt excited
neutral oxygen as the working physical interpretation of the feature.
At the $R\simeq600$ resolution of the ALPY~600 spectrograph, the
individual triplet components cannot be separated, so the present
data do not by themselves provide a unique atomic identification.
Nevertheless, the wavelength coincidence, transit dependence, and
expected oxygen-rich mineral vapor motivate a quantitative
O~{\sc i} interpretation.}

{We test this interpretation with a one-zone NLTE forward model in
which stellar photoionization supplies O$^+$ and recombination with
radiative cascade populates the $3s\,{}^5S^\circ$ lower level. The
model reproduces the observed scale of the equivalent width, line
depth, and instrumentally broadened profile for an oxygen-bearing gas
component extending beyond the narrow condensed-dust layer. We
therefore treat the feature throughout this work as candidate
O~{\sc i} absorption from an escaping mineral-vapor component, while
retaining stellar and terrestrial effects as observational
alternatives that require higher-resolution spectroscopy to separate
definitively.}

{Section~\ref{sec:observations} describes the observations and data
reduction. Section~3 presents the phase-dependent spectroscopic and
photometric results. Section~4 develops the oxygen-bearing
rock-vapor interpretation, the NLTE forward model, and alternative
origins of the candidate feature. Section~5 summarizes our
conclusions.}

\begin{figure}[t]
\centering
\includegraphics[width=0.95\columnwidth]{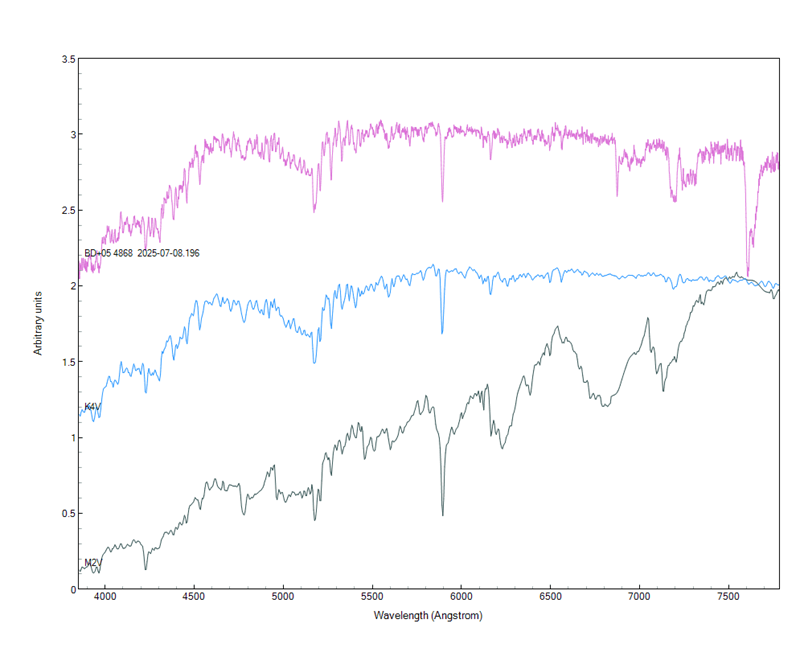}
\caption{{
Representative ALPY~600 spectrum of BD+05~4868~A compared with
K-dwarf and M-dwarf reference spectra from the ALPY~600 spectral
reference library distributed for use with ISIS \citep{Buil2020}.
Because the reference spectra were obtained and prepared for the same
spectrograph class, they provide an appropriate qualitative comparison
at the resolving power of the present observations. The broad
continuum shape and principal photospheric absorption structure of
BD+05~4868~A are consistent with the known K-type primary and differ
substantially from the M-dwarf reference spectrum. The comparison is
used to verify that the optical spectrum is dominated by the K-type
primary rather than as an independent stellar classification or a
quantitative measurement of flux contamination from the known
M-dwarf companion.
}}
\label{fig:spectral_templates}
\end{figure}

\section{Observations and Data Reduction}
\label{sec:observations}

\subsection{Spectroscopic Observations}

\subsubsection{Instrumentation and Observing Strategy}

{Time-resolved optical spectroscopy of BD+05~4868~A was obtained with
the 432~mm PlaneWave CDK17 corrected Dall--Kirkham telescope at the
Leif Everson Observatory in Sturgeon Bay, Wisconsin. The telescope
operates at $f/6.8$ and was equipped with a Shelyak ALPY~600
low-resolution slit spectrograph.}

{The ALPY~600 provides a resolving power of approximately
$R\simeq600$ over the usable wavelength interval
3800--7800~\AA. Thus,}

\begin{equation}
R\equiv\frac{\lambda}{\Delta\lambda}\simeq600,
\qquad
\Delta\lambda_{\rm inst}(7774~{\rm \AA})
\simeq13~{\rm \AA}.
\label{eq:instrument_resolution}
\end{equation}

{The 3.45~\AA\ separation between the outer components of the
O~{\sc i} 7774~\AA\ triplet is therefore substantially smaller than
one instrumental resolution element, and the triplet is unresolved in
the present spectra. No spectral data redward of the established
7800~\AA\ usable limit were included in the analysis.}

{Spectroscopic observations were obtained during multiple predicted
transits between 2025 May 25 and July 28. Observing windows were
selected from the orbital ephemeris of \citet{Hon2025} to sample both
the expected dust-tail occultation and adjacent baseline intervals.
Individual science integrations were typically 600--900~s, with
individual observing sequences spanning approximately 1.5--4.2~hr.
Eight nights yielded usable spectroscopic observations, with
approximately 10--15 exposures acquired per sequence.}

{Assignment to the in-transit and out-of-transit samples was determined
from orbital phase rather than from the appearance of the spectrum
near 7774~\AA. Exposures obtained during the predicted passage of the
extended occulting material across the stellar disk were assigned to
the in-transit sample, while observations sufficiently separated from
the predicted occultation formed the out-of-transit baseline. The
classification was therefore fixed independently of whether a
particular exposure displayed structure near the candidate wavelength.}

{Individual spectra were retained only when they satisfied a common
set of full-spectrum quality criteria. Exposures affected by
substantially reduced continuum signal-to-noise, severe guiding or
slit-centering losses, anomalous broadband throughput, inadequate
sky-background subtraction, unreliable wavelength registration, or
rapid transparency variations were excluded. These criteria were
applied without reference to the strength of the candidate
7774~\AA\ feature.}

\subsubsection{Spectroscopic Reduction and Atmospheric Correction}

{Raw CCD frames were reduced using the Integrated Software for Imagers
and Spectrometers (ISIS; \citealt{Buil2020}). Processing included
dark-current subtraction, flat-field correction, cosmic-ray rejection,
background subtraction, and extraction of the one-dimensional stellar
spectrum. Sky background was estimated from detector regions adjacent
to the stellar trace, and the stellar spectrum was extracted using an
optimal-extraction procedure \citep{Horne1986}.}

{Wavelength calibration was performed using Ne--Ar comparison-lamp
spectra obtained during each observing night. The resulting wavelength
solutions had typical root-mean-square residuals of approximately
0.3~\AA, substantially smaller than the instrumental resolution near
7774~\AA.}

{A nearby comparison star observed at similar airmass and time was used
to characterize large-scale changes in instrumental throughput and
atmospheric transparency. A smooth relative-response correction was
derived while excluding narrow stellar absorption structure from the
response determination. In particular, the comparison spectrum was
not used as a direct divisor in the O~{\sc i} 7774~\AA\ region,
because late-type stellar spectra can themselves contain intrinsic
O~{\sc i} absorption.}

{Telluric absorption was treated on an exposure-by-exposure basis using
a synthetic atmospheric-transmission model following the Molecfit
methodology \citep{Smette2015,Kausch2015}. The atmospheric solution
included O$_2$ and H$_2$O and was constrained using telluric-dominated
regions near the O$_2$ B band around 6870~\AA, H$_2$O structure near
7200--7350~\AA, and the O$_2$ A band near 7600--7700~\AA.
Atmospheric water vapor was allowed to vary between exposures.}

{The 7745--7800~\AA\ interval containing the candidate feature was
excluded from determination of the atmospheric solution. The
high-resolution synthetic transmission spectrum was convolved to the
effective ALPY~600 line-spread function before correction of the
observed spectrum. Regions of very low atmospheric transmission or
poorly corrected telluric structure were excluded from the
differential analysis.}

{Additive night-sky emission was treated separately from
multiplicative telluric absorption. Sky spectra measured adjacent to
the stellar trace were inspected in the vicinity of the terrestrial
O~{\sc i} 7774~\AA\ airglow transition
\citep{Christensen1975}. Exposures showing anomalous residual
background structure were rejected using the same full-spectrum
quality criteria applied elsewhere in the reduction. The telluric and
sky-background procedures were applied identically to the in-transit
and out-of-transit samples.}

\subsubsection{Wavelength Registration and Resolution Matching}

{Before stacking, all accepted spectra were placed on a common
wavelength grid. Residual exposure-to-exposure wavelength offsets were
measured from photospheric absorption features distributed across the
spectrum and outside the candidate O~{\sc i} interval. Each spectrum
was then interpolated onto the common reference grid. The 7774~\AA\
feature itself was not used to determine the wavelength shifts.}

{Small differences in seeing, guiding, and slit illumination can also
change the effective line-spread function between exposures. Because
such differences can generate structured residuals when otherwise
identical stellar spectra are divided, the effective widths of
photospheric absorption features outside the candidate region were
compared between the in-transit and out-of-transit samples.}

{Where required, the spectrum with the narrower effective
line-spread function was broadened to match the broader spectrum.
For approximately Gaussian line-spread functions, the matching kernel
satisfies}

\begin{equation}
\sigma_{\rm ker}
=
\left(
\sigma_{\rm broad}^{2}
-
\sigma_{\rm narrow}^{2}
\right)^{1/2},
\label{eq:lsf_matching}
\end{equation}

{where $\sigma_{\rm broad}$ and $\sigma_{\rm narrow}$ are the measured
Gaussian widths of the broader and narrower profiles, respectively.
No deconvolution was performed. Wavelength registration and
resolution matching were therefore determined from the stellar
spectrum independently of the candidate 7774~\AA\ interval.}

\subsubsection{Construction of the Transmission Spectrum}

{Following atmospheric correction, wavelength registration,
resolution matching, and a common continuum-normalization procedure,
accepted spectra were separated into the phase-defined in-transit and
out-of-transit samples. The final phase-selected stacks contained
$N_{\rm in}=24$ in-transit spectra and
$N_{\rm out}=73$ out-of-transit spectra. Because reliable per-pixel
variance estimates were not used as statistical weights, each accepted
exposure was assigned equal weight. The mean spectrum of each sample
was therefore}

\begin{equation}
\overline{f}(\lambda)
=
\frac{1}{N}
\sum_{i=1}^{N}
f_i(\lambda),
\label{eq:mean_stack}
\end{equation}

{where $f_i(\lambda)$ is the normalized spectrum of exposure $i$ and
$N$ is the number of accepted spectra contributing at that wavelength.
Separate means,
$\overline{f}_{\rm in}(\lambda)$ and
$\overline{f}_{\rm out}(\lambda)$, were constructed for the
in-transit and out-of-transit samples.}

{The exposure-to-exposure scatter within each stack was used to
estimate the uncertainty in its mean spectrum. We then defined the
differential transmission spectrum as}

\begin{equation}
\mathcal{T}(\lambda)
=
\frac{\overline{f}_{\rm in}(\lambda)}
     {\overline{f}_{\rm out}(\lambda)}.
\label{eq:transmission}
\end{equation}

{With this convention,
$\mathcal{T}=1$ indicates no phase-dependent spectral change, whereas
$\mathcal{T}<1$ indicates additional absorption in the in-transit
spectrum.}

{Assuming independent uncertainties in the two mean spectra, the
uncertainty in the transmission spectrum was propagated as}

\begin{equation}
\sigma_{\mathcal T}
=
\mathcal{T}
\left[
\left(
\frac{\sigma_{\rm in}}
     {\overline{f}_{\rm in}}
\right)^2
+
\left(
\frac{\sigma_{\rm out}}
     {\overline{f}_{\rm out}}
\right)^2
\right]^{1/2}.
\label{eq:transmission_uncertainty}
\end{equation}

{The equivalent width of the candidate absorption was measured from
the locally normalized differential transmission spectrum,}

\begin{equation}
W_\lambda =
\int_{\lambda_1}^{\lambda_2}
[1-T_{\rm norm}(\lambda)]\,d\lambda,
\end{equation}

{where the nominal integration interval was
$7754$--$7778$~\AA. A local linear continuum was determined from
the neighboring intervals $7742$--$7752$~\AA\ and
$7780$--$7784$~\AA. The sensitivity of the measurement to reasonable
changes in these boundaries is considered in Section~3.1.}

\subsection{Photometric Observations}

{Complementary time-series photometry was obtained during the
spectroscopic observing campaign to establish the contemporaneous
dust-tail morphology and confirm that BD+05~4868~Ab remained in an
active, asymmetric occultation state. These observations are used as
supporting context for the spectroscopy; the detailed dust-tail
geometry is adopted from the longer-baseline analysis of
\citet{Hon2025} rather than independently derived from the shorter
ground-based sequences.}

\subsubsection{Sierra Remote Observatories}

{}{Photometric observations at Sierra Remote Observatories (SRO) in
California were obtained with a 610~mm PlaneWave CDK24 telescope and
a photometric $V'$ filter with an effective wavelength near 545~nm.
Individual integrations were 60~s, and observing sequences typically
covered approximately 2--3~hr around the predicted occultation.
These data provide the highest signal-to-noise ground-based light
curves in the present sample.}

\subsubsection{Indiana State University}

{Additional photometric observations were obtained at Indiana State
University using a 356~mm Meade LX600 ACF telescope and a Sloan
$R'$ filter with an effective wavelength near 655~nm. Exposure times
were approximately 30--60~s, and individual observing runs covered
roughly 2~hr around the predicted transit. These observations provide
an independent light curve in a redder optical passband.}

\subsubsection{Photometric Reduction}

{Photometric images from both facilities were reduced using
AstroImageJ \citep{Collins2017}. Calibration included dark-current
subtraction and flat-field correction. Aperture photometry was
performed on BD+05~4868~A and a set of nearby comparison stars, with
the aperture radius selected to minimize the scatter of the
differential light curve.}

{For each observing sequence, the relative target flux was calculated
as}

\begin{equation}
F_{\rm rel}(t)
=
\frac{F_{\rm target}(t)}
{\sum_j F_{{\rm comp},j}(t)},
\label{eq:differential_photometry}
\end{equation}

{where $F_{\rm target}$ is the measured target flux and
$F_{{\rm comp},j}$ is the flux of comparison star $j$.
Slow baseline variations associated with airmass or other observing
conditions were removed using low-order trends fitted to the
out-of-transit data, and each light curve was normalized to its
out-of-transit baseline.}

{Observation times were converted to the barycentric time standard
used by the orbital ephemeris. Where BJD$_{\rm TDB}$ is reported, the
conversion follows the timing conventions of \citet{Eastman2010}.}

\begin{figure*}[t]
\centering
\includegraphics[width=0.7\textwidth]
{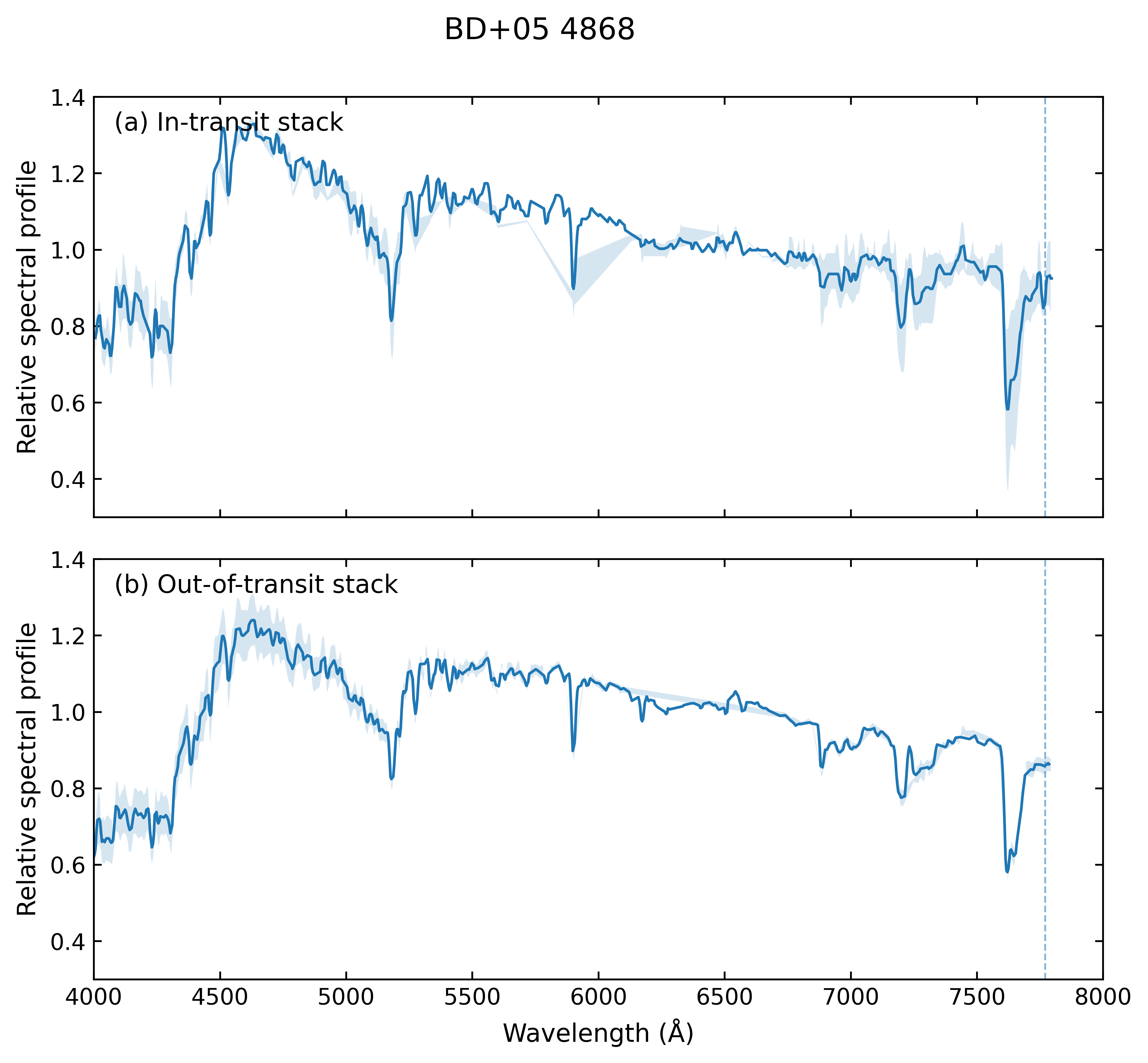}
\caption{{
Mean in-transit (top) and out-of-transit (bottom) ALPY~600 spectra
of BD+05~4868~A after wavelength registration, continuum treatment,
and matching of the effective spectral resolution. The shaded regions
show the $1\sigma$ uncertainty in each mean stack estimated from the
exposure-to-exposure scatter. The spectra are shown over the usable
3800--7800~\AA\ wavelength range. The vertical dashed line marks the
region of the O~{\sc i} 7774~\AA\ triplet and is included only as a
wavelength reference. Across most of the optical spectrum the two
phase-selected stacks show similar broad photospheric structure, while
the principal candidate phase-dependent difference occurs near the
red end of the spectrum.
}}
\label{fig:stacked_spectra}
\end{figure*}

\begin{figure*}[t]
\centering
\includegraphics[width=0.70\textwidth]
{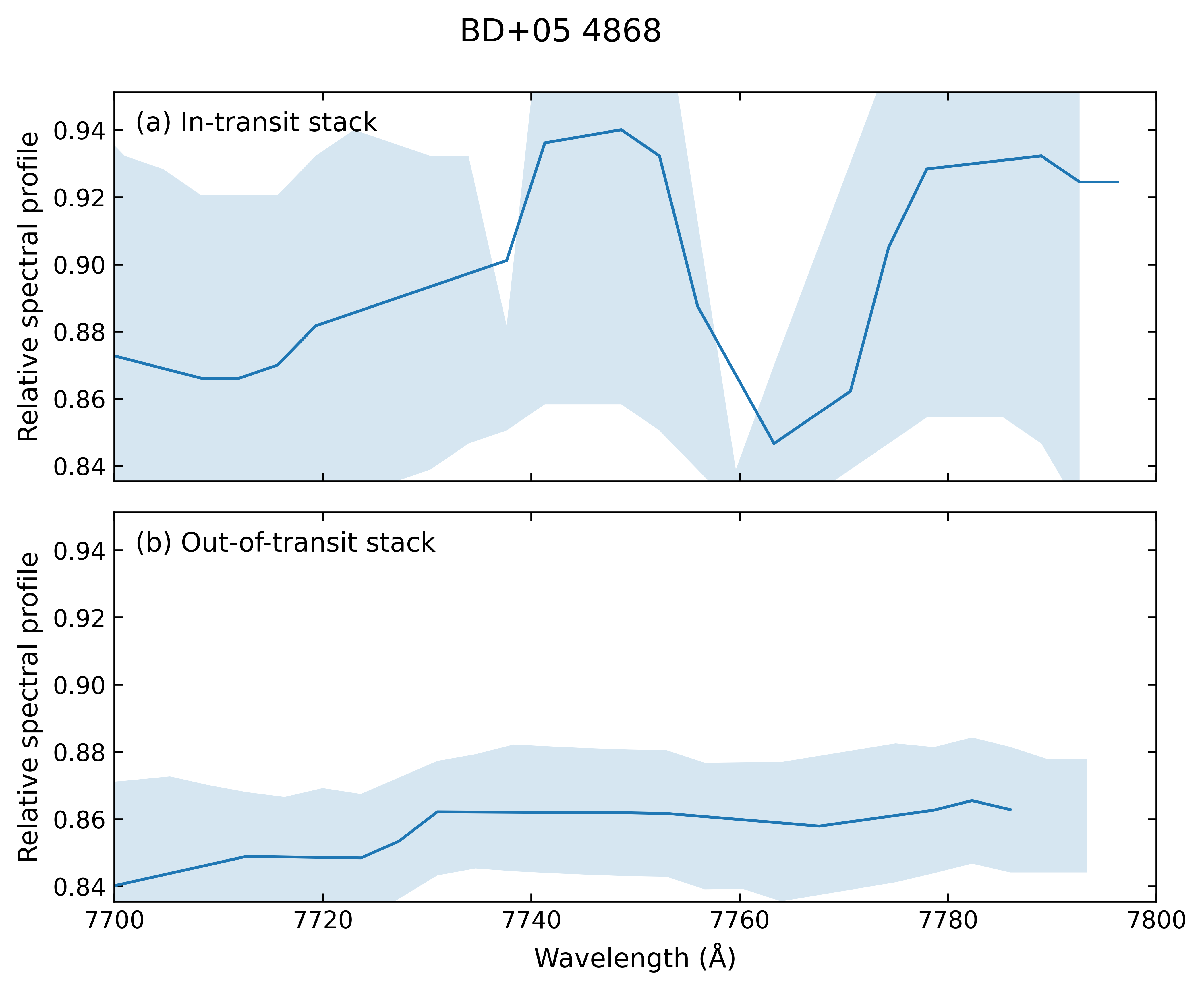}
\caption{{
Red-end comparison of the stacked ALPY~600 spectra of
BD+05~4868~A over 7700--7800~\AA. (a) Mean in-transit spectrum
constructed from exposures obtained during the predicted dust-tail
occultation. (b) Mean out-of-transit comparison spectrum constructed
from the phase-selected baseline observations. The shaded regions show the $1\sigma$ uncertainty in each mean stack
estimated from the exposure-to-exposure scatter. An additional
absorption-like depression is present in the in-transit spectrum near
the wavelength of the O~{\sc i} 7774~\AA\ triplet relative to the
out-of-transit spectrum. The three O~{\sc i} components span only
3.45~\AA\ and are unresolved within the approximately 13~\AA\
ALPY~600 instrumental profile; the present spectra therefore show a
single blended candidate feature rather than the individual triplet
components.
}}
\label{fig:oi_region}
\end{figure*}

\section{Results}
\label{sec:results}

\begin{figure*}[t]
\centering
\includegraphics[width=0.7\textwidth]
{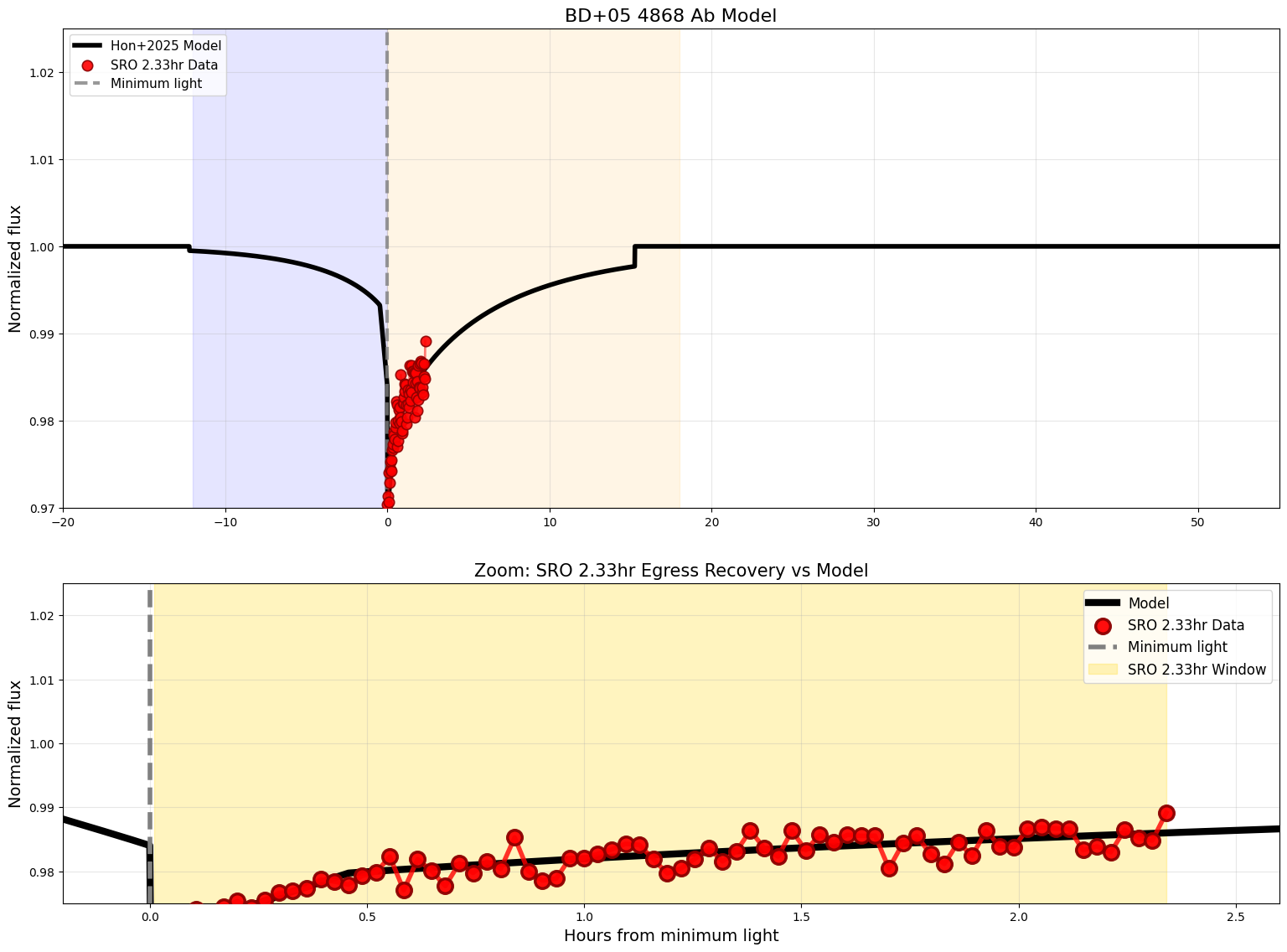}
\caption{{
Comparison of the SRO photometry with the published dust-tail
morphology of BD+05~4868~Ab. Top: mean transit profile calculated
from the generalized-Gaussian prescription and published parameters
of \citet{Hon2025}, with the representative SRO $V'$-band sequence
overplotted. The vertical dashed line denotes minimum light.
Bottom: expanded view of the approximately 2.33~hr SRO observing
interval following minimum light. The data show a gradual recovery
broadly consistent with the extended trailing-tail morphology
predicted by the published dust distribution. The model curve was
calculated from the published Hon et al. parameters and was not
refitted to the present observations.
}}
\label{fig:photometry_hon}
\end{figure*}

\subsection{Transit-Correlated Absorption near 7774~\AA}
\label{sec:results_spectra}

{The reduced ALPY~600 spectrum of BD+05~4868~A exhibits the broad
continuum shape and blended photospheric absorption structure expected
for the K-type primary. A representative spectrum is compared with
K- and M-dwarf templates in Figure~\ref{fig:spectral_templates}.
The target spectrum is substantially more consistent with the K-dwarf
template, supporting the conclusion that the primary dominates the
observed optical spectrum.}

{The mean in-transit and out-of-transit spectra are shown over the full
usable 3800--7800~\AA\ range in
Figure~\ref{fig:stacked_spectra}. Across most of the spectrum, the
two phase-selected stacks reproduce the same broad photospheric
structure after wavelength registration, continuum treatment, and
matching of their effective spectral resolution. The principal
phase-dependent difference occurs near the red end of the spectrum,
where the in-transit stack contains an additional absorption-like
depression near 7774~\AA.}

{Figure~\ref{fig:oi_region} shows this region in greater detail.
The additional absorption is present in the in-transit stack near the
laboratory wavelengths of the O~{\sc i} triplet at 7771.94, 7774.17,
and 7775.39~\AA, while a comparable depression is not present in the
out-of-transit stack. Because the complete triplet spans only
3.45~\AA, substantially less than the approximately 13~\AA\
instrumental resolution established in
Section~\ref{sec:observations}, the individual components cannot be
distinguished. The observed feature is therefore treated as a single
unresolved absorption feature near the O~{\sc i} 7774~\AA\ triplet.}

{The differential transmission spectrum defined in
Equation~\ref{eq:transmission} reaches a maximum absorption depth of
approximately 9\% in the candidate region. Its apparent width is
approximately 13--16~\AA, consistent with an unresolved feature
dominated by the ALPY~600 line-spread function. The present data
therefore constrain the integrated absorption strength but not the
intrinsic velocity profile or the relative strengths of the three
triplet components.}

{Using the locally normalized transmission spectrum and integrating
over the nominal 7754--7778~\AA\ interval gives
$W_\lambda=1.40$~\AA. Varying the integration boundaries over
reasonable intervals spanning the unresolved feature produces values
of approximately $1.32$--$1.42$~\AA. To conservatively account for
the additional sensitivity to local continuum placement, we adopt
$W_\lambda\simeq1.4\pm0.2$~\AA. The quoted uncertainty is a
methodological uncertainty intended to encompass reasonable continuum
and integration choices rather than a formal Gaussian detection
significance.}

{A profile constrained to the instrumental resolution gives a line
strength consistent with the direct equivalent-width measurement.
Allowing the width to vary produces a somewhat broader profile, but
the difference is insufficient to establish intrinsic broadening
beyond the instrumental response. Likewise, no physically meaningful
velocity centroid can be assigned to an unresolved feature this close
to the red limit of the usable spectral range. We therefore identify
the observational result as a transit-correlated, unresolved
absorption feature near the O~{\sc i} 7774~\AA\ triplet.}

\subsubsection{Checks on the Candidate Feature}
\label{sec:spectral_checks}

{The candidate interval lies redward of the strongest portion of the
terrestrial O$_2$ A band. The atmospheric correction described in
Section~\ref{sec:observations} was constrained from independent
telluric regions, with the 7745--7800~\AA\ interval excluded from
determination of the atmospheric solution. Thus, the fitted
atmospheric model was not optimized using the candidate absorption
itself.}

{Additive sky emission was treated independently from telluric
absorption using background measured adjacent to the stellar trace.
Wavelength registration and effective line-spread-function matching
were likewise determined from photospheric features outside the
candidate interval. The absorption-like difference near 7774~\AA\
remains after these corrections and is not accompanied by a comparable
phase-dependent residual across most of the optical spectrum.}

{Taken together, these checks support treating the 7774~\AA\ feature
as an astrophysical candidate associated with the transit-selected
spectra. Its identification with planetary O~{\sc i}, however, also
depends on stellar line-profile effects that cannot be separated at
the present spectral resolution. Those alternatives are considered in
Section~\ref{sec:discussion}.}

\subsection{Ground-Based Dust-Tail Photometry}
\label{sec:results_photometry}

{The ground-based photometry confirms that BD+05~4868~Ab continued to
exhibit a broad, asymmetric occultation during the observing campaign.
The SRO light curves show a structured minimum followed by a gradual
recovery toward the out-of-transit baseline rather than the sharp,
symmetric egress expected from a compact opaque planet. Independent
ISU observations likewise show variability associated with the
predicted occultation, although at lower photometric precision.}

{Figure~\ref{fig:photometry_hon} compares a representative SRO
sequence with the mean dust-tail morphology calculated from the
generalized-Gaussian prescription and published parameters of
\citet{Hon2025}. The SRO observations sample approximately 2.33~hr
of the post-minimum-light recovery and follow the broad rising trend
expected from extinction by an extended trailing tail. The published
model is shown for morphological comparison and is not refitted to the
present ground-based data.}

{For the mean transit profile, \citet{Hon2025} obtained characteristic
leading- and trailing-tail scale lengths of approximately
$x_{\rm f}=0.58R_\star$ and $x_{\rm b}=3.6R_\star$, respectively,
with a generalized-Gaussian exponent $g\simeq0.85$. The much larger
trailing scale naturally produces a prolonged post-minimum recovery
similar to that sampled by the SRO observations. The detailed
point-to-point structure of the ground-based light curve does not
exactly follow the mean model, which is expected given the strong
transit-to-transit variability already established for this system.}

{We therefore use the photometry to establish that the system remained
in an active, asymmetric dust-producing state during the spectroscopic
campaign. The shorter ground-based sequences are not used to derive an
independent dust-tail length or mass-loss rate; the quantitative dust
geometry employed in the physical interpretation is taken from the
longer-baseline analysis of \citet{Hon2025}.}

\section{Discussion}
\label{sec:discussion}

{The phase-dependent absorption near 7774~\AA\ is consistent in
wavelength, strength, and instrumental width with the unresolved
O~{\sc i} triplet. We therefore adopt excited neutral oxygen in the
escaping mineral-vapor flow as the working physical interpretation of
the feature. The principal questions are whether sufficient oxygen is
expected in the outflow, whether the excited lower state of the
7774~\AA\ transition can be populated under the conditions of
BD+05~4868~Ab, and whether the observed absorption can be reproduced
quantitatively without requiring extreme gas properties.}

\subsection{An Oxygen-Bearing Rock-Vapor Outflow}
\label{sec:oxygen_vapor}

{An oxygen-bearing gaseous component is a natural consequence of rocky
material being heated to vaporization temperatures. Thermochemical
calculations for volatile-poor silicate material predict gas-phase
O$_2$, O, SiO, Na, Mg, Fe, and related species, with the detailed
abundances determined by bulk composition, temperature, pressure, and
fractional vaporization
\citep{SchaeferFegley2009,SchaeferLoddersFegley2012,
Costa2017,ItoIkoma2021}. The escaping dust and gas should therefore
be regarded as different phases of the same mineral-vapor outflow
rather than as physically unrelated components.}

{The chemical pathway connecting oxygen-bearing vapor to excited
neutral oxygen was summarized in Equation~\ref{eq:oxygen_pathway}.
Stellar irradiation can photodissociate O$_2$, photoionize neutral
oxygen, and maintain a modest population of O$^+$ that subsequently
recombines into excited states of O~{\sc i}. Radiative cascades through
the quintet system then provide a direct route to the
$3s\,{}^5S^\circ_2$ lower level of the 7774~\AA\ triplet
\citep{MaurerMazzali2010,Nahar2022}.}

{The high material-loss rate of BD+05~4868~Ab strengthens the
connection between the gaseous and condensed components of the
outflow. \citet{Hon2025} infer a dust mass-loss rate of order
$2\times10^{12}~{\rm g\,s^{-1}}$ and show that vaporization fluxes
from rock-forming material can approach those required to supply the
observed dusty effluent. The same mineral-vapor source capable of
producing the dust can therefore naturally leave a residual gaseous
component, including atomic oxygen. Their dust simulations further
produce grains predominantly in the $\sim1$--$10~\mu{\rm m}$ range
and an extended trailing tail, supporting a picture in which freshly
vaporized rocky material supplies both the observed grains and the
gas from which they form.}

{The gas need not remain geometrically identical to the condensed dust.
Atoms and ions may expand beyond the region in which grains remain
concentrated, while condensation and sublimation continuously exchange
material between the two phases. The dust-tail geometry therefore
provides a useful reference scale for the absorbing material but does
not uniquely determine the spatial distribution of atomic oxygen.}

\subsection{NLTE Excitation and Forward Modeling of the O~{\sc i}
7774~\AA\ Triplet}
\label{sec:nlte}

\begin{figure}[t]
\centering
\includegraphics[width=\columnwidth]
{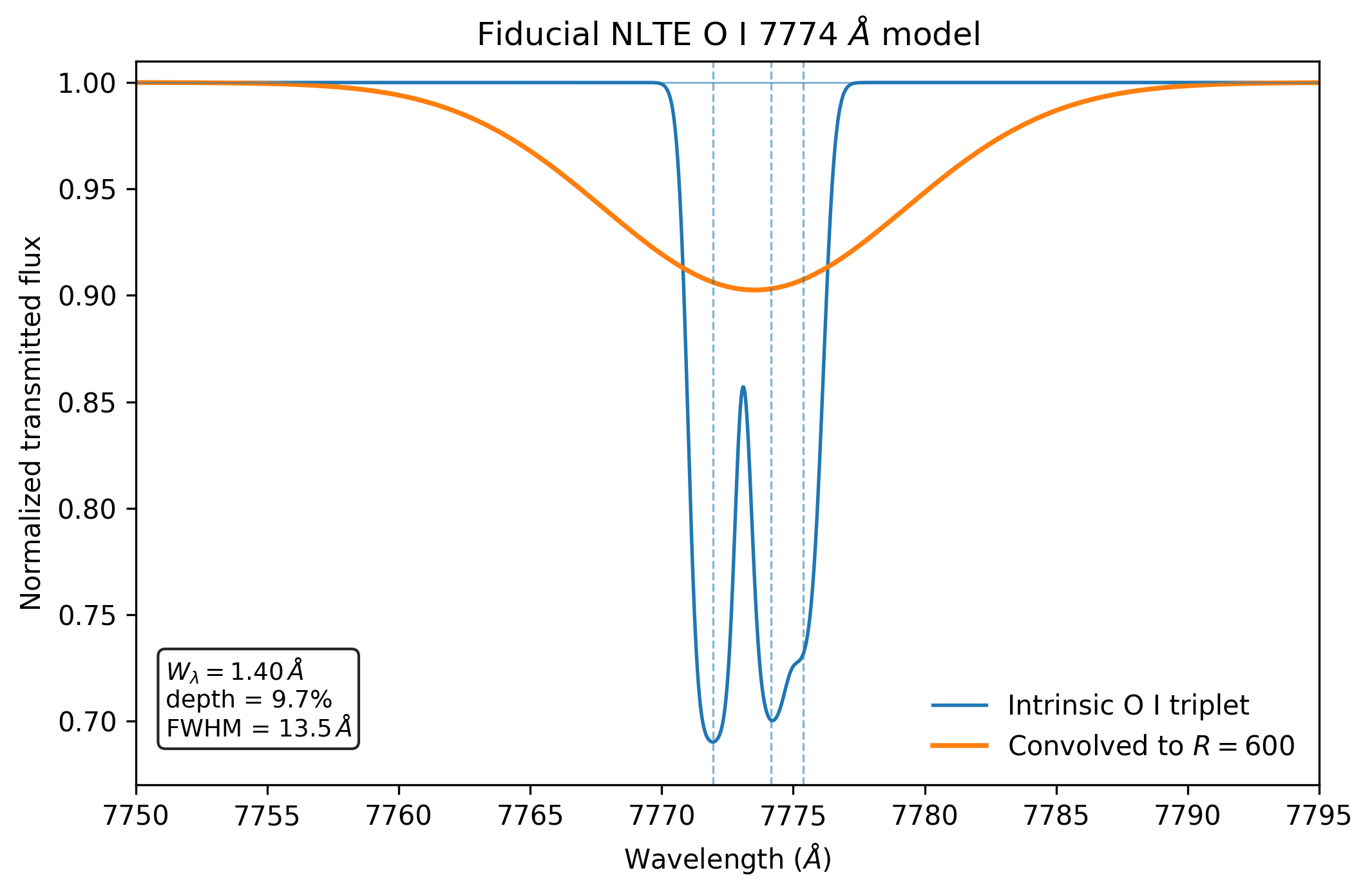}
\caption{{
Representative forward NLTE model of the O~{\sc i}
7774~\AA\ triplet. The intrinsic profile includes absorption from
all three triplet components, while the broader profile shows the
same model after convolution to the ALPY~600 resolving power of
$R\simeq600$. The representative solution reproduces
$W_\lambda=1.40$~\AA\ and yields an apparent differential depth of
approximately 9.7\% and an instrumentally broadened FWHM of
approximately 13.5~\AA. Dashed lines mark the laboratory wavelengths
of the individual O~{\sc i} triplet components. The agreement in
equivalent width, depth, and instrumental width demonstrates that the
NLTE oxygen-tail model can reproduce the observed scale of the
unresolved candidate feature; the adopted parameters represent one
solution within the broader model family.
}}
\label{fig:nlte_profile}
\end{figure}

{The central excitation requirement follows directly from
Equation~\ref{eq:boltzmann_intro}. The
$3s\,{}^5S^\circ_2$ lower state lies 9.146~eV above the ground state,
and its LTE population at the reported $\sim1820$~K substellar
temperature is suppressed by a Boltzmann factor of order
$5\times10^{-26}$. Thermal excitation from ground-state O~{\sc i}
therefore cannot account for measurable 7774~\AA\ absorption.
The line requires an NLTE population mechanism.}

{We therefore adopt a one-zone NLTE treatment in which the absorbing
region is represented by characteristic values of the electron
temperature, gas density, ionization state, and projected geometry.
The calculation is intended as a forward model: it tests whether an
oxygen-bearing outflow with reasonable physical properties can
reproduce the observed absorption, rather than attempting a unique
spatially resolved retrieval of the tail.}

{We model the lower level as being supplied primarily by radiative
recombination of O$^+$ followed by cascade through the excited quintet
system. In the regime relevant to the forward model, the dominant
steady-state balance can be written}

\begin{equation}
n_{5S}
\simeq
\frac{
\alpha_{\rm eff}(T_e)\,n_e n_{{\rm O}^{+}}
}{
A_{1355}\beta_{\rm esc}+\Gamma_{5S}
},
\label{eq:nlte_population}
\end{equation}

{Here, $\alpha_{\rm eff}(T_e)$ is the effective recombination
coefficient feeding the excited quintet system, $n_e$ is the electron
density, and $n_{{\rm O}^{+}}$ is the O~{\sc ii} number density.
The numerator therefore represents the rate at which recombination
followed by radiative cascade populates the $3s\,{}^5S^\circ$ state.
The denominator describes the principal loss processes from that
state. The term $A_{1355}\beta_{\rm esc}$ is the effective radiative
loss rate through the weak transition near 1355~\AA, while
$\Gamma_{5S}$ represents photoionization from the excited state.
The escape probability is approximated by}

\begin{equation}
\beta_{\rm esc}
=
\frac{1-e^{-\tau_{1355}}}{\tau_{1355}}.
\label{eq:escape_probability}
\end{equation}

{When the 1355~\AA\ transition becomes optically thick,
$\beta_{\rm esc}<1$, because photons emitted through this transition
are less likely to escape the absorbing region directly. The effective
radiative depopulation rate of the $3s\,{}^5S^\circ$ state is
therefore reduced, allowing a larger steady-state population to remain
available for absorption in the 7774~\AA\ triplet. Radiative trapping
can consequently make the recombination-driven population of this
otherwise sparsely occupied excited state substantially more effective.}

{The absorption calculation treats all three triplet components with
their full optical depths and a projected covering fraction $C_f$.
The corresponding model transmission is}

\begin{equation}
\mathcal{T}_{\rm mod}(\lambda)
=
1-C_f\left[1-e^{-\tau_\lambda}\right].
\label{eq:nlte_transmission}
\end{equation}

{The strongest portion of the modeled triplet is optically thick, with
a maximum optical depth of approximately $\tau_{7774}\sim4$.
At this optical depth, the line-center transmission through a covered
sightline is only $e^{-4}\simeq0.02$, so the intrinsic line core is
effectively saturated rather than optically thin. We therefore
calculate the transmitted spectrum using the full
$1-e^{-\tau_\lambda}$ dependence in Equation~\ref{eq:nlte_transmission}
rather than the optically thin approximation
$1-e^{-\tau_\lambda}\simeq\tau_\lambda$. The disk-integrated
absorption remains considerably shallower because the gas covers only
a fraction of the projected stellar disk.}

\subsubsection{Constraints from the Observed Feature}

{We constructed a one-zone NLTE forward model constrained by the
observed $W_\lambda\simeq1.4$~\AA\ equivalent width. Starting from
an assumed gas geometry and mass-loss rate, the calculation determines
a characteristic gas density and evaluates the
O~{\sc i}/O~{\sc ii} ionization balance. Recombination and radiative
cascade then determine the population of the
$3s\,{}^5S^\circ$ lower state, including the effect of trapping in
the 1355~\AA\ transition. The model calculates the full optical depths
of all three O~{\sc i} 7774~\AA\ components, applies the projected
stellar covering fraction, and finally convolves the resulting
transmission spectrum with the ALPY~600 instrumental response for
direct comparison with the observations.}

{Table~\ref{tab:nlte_parameters} summarizes the principal parameters
and calculated quantities for the representative NLTE solution.}

{As a representative solution, we adopt a projected atomic-gas
half-height of $h_{\rm gas}=0.25R_\star$, corresponding in the
adopted strip geometry to a stellar covering fraction of
$C_f\simeq0.315$. Here, $h_{\rm gas}$ describes the projected
extent of the absorbing tail rather than a hydrostatic atmospheric
scale height.}

{We further adopt a representative electron temperature of
$T_e=7500$~K and an ionizing luminosity of
$L_{\rm EUV}/L_{\rm bol}=3\times10^{-6}$. The electron temperature
describes the tenuous irradiated gas and is distinct from the
$\sim1820$~K characteristic temperature associated with the planetary
surface. The adopted EUV luminosity is likewise a model assumption
rather than a direct measurement of the host star.}

{For these parameters, reproducing the measured equivalent width
requires
$\dot M_{\rm gas}/\dot M_{\rm dust}\simeq5.1$. Combined with the
dust mass-loss estimate
$\dot M_{\rm dust}\sim2\times10^{12}~{\rm g\,s^{-1}}$ from
\citet{Hon2025}, the representative solution corresponds to
$\dot M_{\rm gas}\sim1\times10^{13}~{\rm g\,s^{-1}}$. This value
is a model-dependent requirement of the fiducial solution rather than
a direct measurement of the gas mass-loss rate.}

{The corresponding total oxygen column along the modeled line of sight
is of order $2\times10^{20}~{\rm cm^{-2}}$, whereas only
$N(3s\,{}^5S^\circ)\simeq2.1\times10^{13}~{\rm cm^{-2}}$ is
required in the excited lower level of the optical triplet. Thus, only
approximately $10^{-7}$ of the total oxygen column---roughly one oxygen
atom in $10^7$---must occupy the $3s\,{}^5S^\circ$ state at a given
time. The representative ionization balance remains predominantly
neutral, with only of order several percent of the oxygen present as
O~{\sc ii}. The required absorption therefore does not demand either
a highly ionized oxygen tail or a large excited-state fraction.}

{The weak 1355~\AA\ transition is itself optically thick in the
representative solution, with $\tau_{1355}$ of order $10^2$.
Equation~\ref{eq:escape_probability} then gives an escape probability
of order $\beta_{\rm esc}\sim10^{-2}$. The effective radiative
depopulation rate of the $3s\,{}^5S^\circ$ state is therefore reduced
by roughly two orders of magnitude relative to the freely escaping
case, helping the recombination-driven population required by the
optical triplet to persist.}

{After convolution with the $R\simeq600$ ALPY~600 instrumental
response, the same model produces an apparent differential depth of
approximately 9.7\% and a full width at half maximum (FWHM) of
approximately 13.5~\AA, in close agreement with the observed scale of
the unresolved feature. The FWHM is the total wavelength width of the
absorption profile measured at one-half of its maximum depth. Its value
is very close to the approximately 13~\AA\ instrumental resolution
near 7774~\AA, showing that the modeled feature would appear
essentially unresolved in the present spectra. The agreement in
equivalent width, depth, and instrumentally broadened width demonstrates
that an oxygen-bearing NLTE outflow can quantitatively reproduce the
observed 7774~\AA\ absorption.}

{The characteristic bulk flow speed and the velocity width of the
absorption profile represent different physical quantities in the
model. The adopted $v_{\rm flow}=0.4~{\rm km\,s^{-1}}$ enters the
mass-continuity estimate and therefore determines the characteristic
gas density associated with a given mass-loss rate; schematically,
$\rho\propto\dot M_{\rm gas}/v_{\rm flow}$ for a fixed flow area.
A slower bulk flow therefore produces a larger density for the same
mass-loss rate, while a faster flow produces a lower density.}

{In contrast, the absorption calculation uses an effective macroscopic
velocity scale of approximately $27~{\rm km\,s^{-1}}$ to parameterize
the unresolved spread of line-of-sight velocities within the extended
absorbing structure. This parameter controls the intrinsic spectral
width of the modeled triplet and should not be interpreted as a
$27~{\rm km\,s^{-1}}$ launch speed from the planetary surface.}

\begin{table*}[t]
\centering
\caption{{Representative One-Zone NLTE O~{\sc i} Model}}
\label{tab:nlte_parameters}

\begin{tabular}{lll}
\hline
\hline
Quantity & Symbol & Value \\
\hline

Electron temperature
& $T_e$
& $7500~{\rm K}$ \\

Gas-tail half-height
& $h_{\rm gas}$
& $0.25\,R_\star$ \\

Projected covering fraction
& $C_f$
& $0.315$ \\

Bulk flow speed
& $v_{\rm flow}$
& $0.40~{\rm km\,s^{-1}}$ \\

Macroscopic velocity scale
& $b_{\rm macro}$
& $27.45~{\rm km\,s^{-1}}$ \\

Oxygen mass fraction
& $X_{\rm O}$
& $0.40$ \\

Ionizing luminosity fraction
& $L_{\rm EUV}/L_{\rm bol}$
& $3.0\times10^{-6}$ \\

Dust mass-loss rate
& $\dot{M}_{\rm dust}$
& $2.0\times10^{12}~{\rm g\,s^{-1}}$ \\

Instrumental resolving power
& $R$
& $600$ \\

\hline

Gas-to-dust mass-loss ratio
& $\dot{M}_{\rm gas}/\dot{M}_{\rm dust}$
& $5.148$ \\

Gas mass-loss rate
& $\dot{M}_{\rm gas}$
& $1.03\times10^{13}~{\rm g\,s^{-1}}$ \\

Total oxygen column
& $N_{\rm O}$
& $2.04\times10^{20}~{\rm cm^{-2}}$ \\

O~{\sc ii} fraction
& $n_{\rm OII}/n_{\rm O}$
& $0.071$ \\

Excited O~{\sc i} column
& $N(3s\,{}^5S^\circ)$
& $2.07\times10^{13}~{\rm cm^{-2}}$ \\

Excited-state fraction
& $N(3s\,{}^5S^\circ)/N_{\rm O}$
& $1.01\times10^{-7}$ \\

1355~\AA\ optical depth
& $\tau_{1355}$
& $92.9$ \\

1355~\AA\ escape probability
& $\beta_{\rm esc}$
& $0.0108$ \\

Maximum 7774~\AA\ optical depth
& $\tau_{7774,\rm max}$
& $4.11$ \\

Model equivalent width
& $W_\lambda$
& $1.40~\AA$ \\

Convolved differential depth
& $d_{\rm conv}$
& $9.75\%$ \\

Convolved FWHM
& ${\rm FWHM}_{\rm conv}$
& $13.51~\AA$ \\

\hline
\end{tabular}

\vspace{0.5em}

{\parbox{0.95\textwidth}{
\footnotesize
The listed quantities correspond to the representative one-zone
NLTE solution used in the forward-model calculations. The electron
temperature, gas geometry, oxygen mass fraction, ionizing luminosity,
flow speed, and macroscopic velocity scale are fiducial model
parameters. The projected covering fraction follows from the adopted
geometry, while the gas-to-dust mass-loss ratio is solved for by
requiring the model to reproduce $W_\lambda=1.40$~\AA. The remaining
quantities describe the resulting representative solution and should
not be interpreted as a unique retrieval of the outflow properties.
}}

\end{table*}

{The strongest constraint emerging from the model is geometrical.
A gas distribution confined to the approximately $0.05R_\star$
vertical half-height of the simulated trailing dust layer covers only
a small fraction of the projected stellar disk. Because the modeled
O~{\sc i} triplet is already optically thick near line center,
increasing the oxygen column within such a narrow projected region
produces progressively smaller increases in absorption strength.
Reproducing the measured equivalent width in this geometry therefore
requires very large gas columns and correspondingly large
gas-to-dust mass-loss ratios.}

{As the projected vertical extent of the atomic component increases,
its stellar covering fraction also increases. The larger covering area
allows the measured equivalent width to be reproduced with
substantially less gas. Atomic-gas half-heights of approximately
$0.2$--$0.4R_\star$ therefore permit considerably less extreme
solutions than a gas distribution confined to the narrow condensed-dust
layer.}

\begin{figure}[t]
\centering
\includegraphics[width=\columnwidth]
{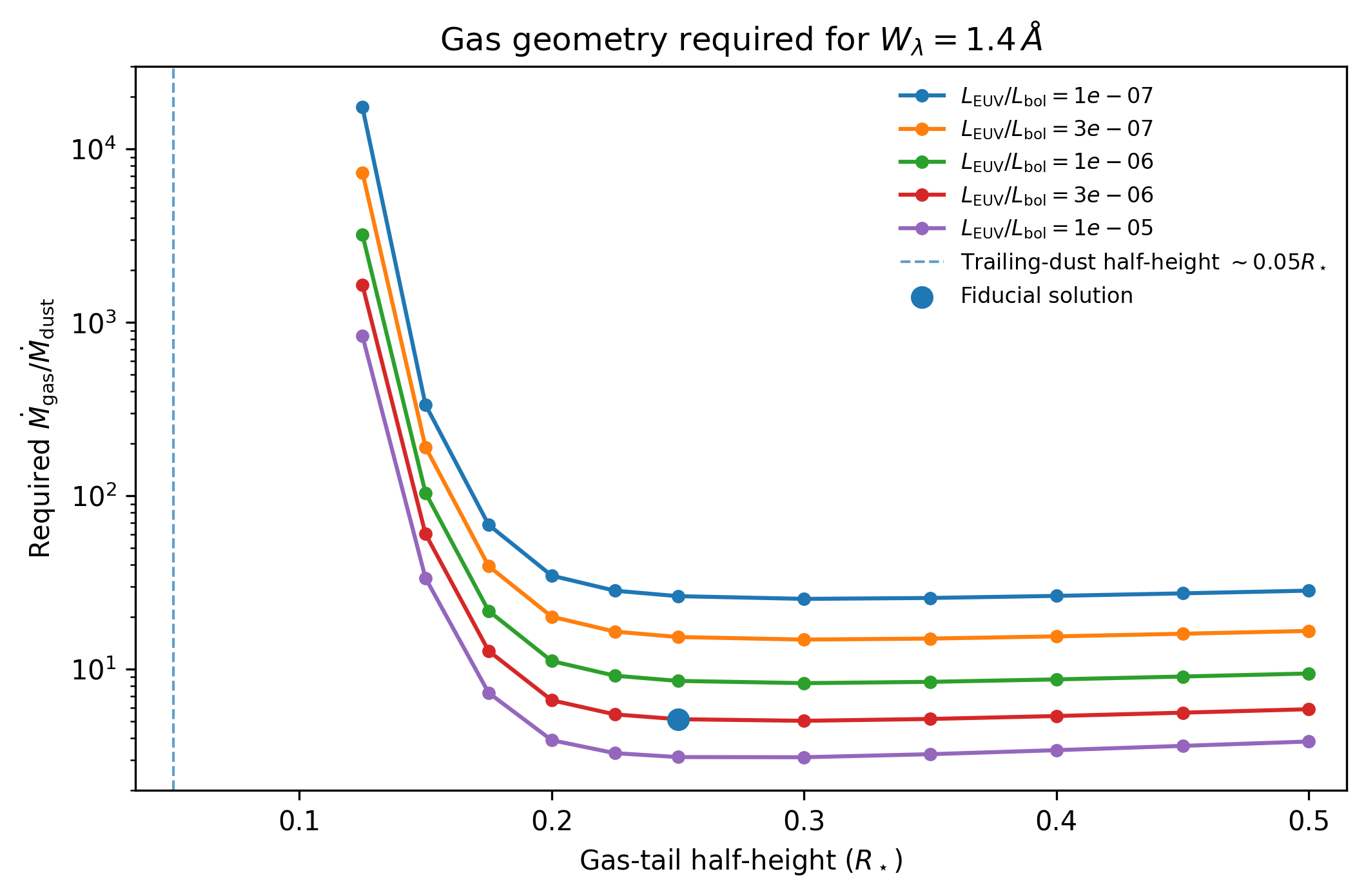}
\caption{{
Dependence of the one-zone NLTE solution on projected gas-tail
extent and stellar EUV irradiation. Curves show the gas-to-dust
mass-loss ratio required to reproduce
$W_\lambda=1.4$~\AA\ as a function of gas-tail half-height for
representative values of $L_{\rm EUV}/L_{\rm bol}$. The dashed
vertical line marks the approximately $0.05R_\star$ half-height of
the simulated trailing dust distribution of \citet{Hon2025} for
comparison. The highlighted representative solution has
$h_{\rm gas}=0.25R_\star$ and
$L_{\rm EUV}/L_{\rm bol}=3\times10^{-6}$, yielding
$\dot M_{\rm gas}/\dot M_{\rm dust}\simeq5.1$. Within the adopted
geometry, moderate gas-to-dust ratios become possible once the
projected atomic component extends beyond the narrow condensed-dust
layer. The family of curves quantifies the degeneracy among gas
extent, ionizing irradiation, and the gas mass-loss rate.
}}
\label{fig:nlte_geometry}
\end{figure}

{Figure~\ref{fig:nlte_geometry} quantifies this behavior. For the
explored range
$L_{\rm EUV}/L_{\rm bol}\sim10^{-6}$--$10^{-5}$,
solutions near the fiducial extended-gas geometry require
$\dot M_{\rm gas}/\dot M_{\rm dust}\sim3$--$9$. These ratios are
model-dependent rather than direct measurements, but they demonstrate
that the observed equivalent width can be reproduced without an
extreme gas content if the atomic component extends beyond the
condensed dust. The family of solutions also illustrates the
degeneracy among projected gas extent, ionizing irradiation, and gas
mass-loss rate.}

{The O~{\sc i} solution is therefore not restricted to a single finely
tuned combination of parameters. Variations in the assumed EUV
irradiation, oxygen abundance, recombination efficiency, projected gas
geometry, or characteristic flow speed shift the inferred
gas-to-dust mass-loss ratio, but an extended oxygen-bearing component
continues to reproduce the measured absorption over a substantial
region of parameter space. We therefore interpret the NLTE calculation
as quantitative support for the O~{\sc i} identification rather than
as a unique retrieval of the gas properties.}

\subsection{Alternative Origins of the 7774~\AA\ Feature}
\label{sec:alternatives}

{The O~{\sc i} interpretation provides a chemically natural and
quantitatively self-consistent explanation of the observed feature,
but the present resolving power does not separate planetary absorption
from all stellar and terrestrial contributions. The most important
remaining alternative is the intrinsic stellar O~{\sc i} triplet
itself.}

{The O~{\sc i} 7774~\AA\ lines are present in the photosphere of
BD+05~4868~A and exhibit non-LTE center-to-limb variations in late-type
stars \citep{Takeda2003,Amarsi2018,Bergemann2021}. During an
occultation, the dust tail removes light from only part of the stellar
disk. If the local O~{\sc i} profile in the occulted region differs
from the disk-integrated stellar spectrum, the in-transit/out-of-transit
ratio can contain residual line structure even in the absence of
planetary oxygen
\citep{Czesla2015,BorsaZannoni2018,Reiners2023}. At
$R\simeq600$, the stellar triplet and any additional planetary
component are blended into the same instrumental profile. Stellar
line-profile variations therefore remain the principal astrophysical
alternative to the escaping-oxygen interpretation.}

{Terrestrial contamination is less directly coincident with the
candidate wavelength. The strongest O$_2$ A-band absorption lies
blueward of 7774~\AA, while variable H$_2$O can produce weaker
structure in the red optical. The candidate interval was excluded from
determination of the atmospheric-transmission solution so that the
absorption near 7774~\AA\ could not influence its own telluric
correction. The same atmospheric procedure was then applied to both
the in-transit and out-of-transit samples.}

{Multiplicative telluric absorption and additive night-sky emission
were treated separately. Molecular absorption by Earth's atmosphere
attenuates the transmitted stellar spectrum, whereas terrestrial
O~{\sc i} airglow contributes additional photons to the extracted
spectrum. The latter was therefore treated through local
sky-background subtraction using detector regions adjacent to the
stellar trace rather than through the atmospheric-transmission model.
Wavelength registration and effective line-spread-function matching
were likewise determined from photospheric features outside the
candidate interval.}

{The persistence of the 7774~\AA\ depression after these independently
determined corrections makes a simple telluric, sky-background, or
resolution-mismatch origin less favored, although residual systematic
effects cannot be excluded completely.}

{We therefore retain the designation ``candidate O~{\sc i}'' because
the low-resolution spectra cannot independently separate the stellar
and planetary contributions. Within that limitation, the combination
of phase dependence, wavelength coincidence, rock-vapor chemistry, and
the quantitative NLTE forward model makes escaping O~{\sc i} our
preferred physical interpretation of the feature.}

\subsection{Relation to Other Disintegrating Rocky Planets}
\label{sec:comparison}

{BD+05~4868~Ab extends a small group of systems in which catastrophic
rocky-planet mass loss can be studied directly. Kepler-1520~b,
KOI-2700~b, and K2-22~b exhibit the same broad phenomenology of
variable transit depths and asymmetric occultations produced by
escaping dusty material
\citep{Rappaport2012,Rappaport2014,SanchisOjeda2015}.
Optical searches for atomic species such as Na~{\sc i} and
Ca~{\sc ii} in these systems have generally resulted in upper limits
rather than secure detections
\citep{Gaidos2019,RiddenHarper2019}.}

{Infrared observations provide a complementary view of the condensed
phase. JWST/MIRI spectroscopy of K2-22~b disfavors a featureless
iron-dominated dust spectrum and is broadly consistent with
silicate-rich material, with additional structure potentially
associated with gaseous species \citep{Tusay2025}. Such observations
probe mineral and molecular signatures, whereas the 7774~\AA\
feature considered here probes a specific excited state of atomic
oxygen.}

{If the O~{\sc i} interpretation is confirmed, BD+05~4868~Ab would
therefore connect the condensed and atomic phases of a disintegrating
rocky-planet outflow: silicate vapor supplies both the dust observed
photometrically and the oxygen-bearing gas traced spectroscopically.
The comparatively bright K-dwarf host and percent-level occultations
make this system particularly favorable for such phase-resolved
spectroscopy \citep{Hon2025}.}

{We do not require BD+05~4868~Ab to occupy a unique evolutionary stage
relative to the other disintegrating systems. Differences in bulk
composition, irradiation, planetary mass, condensation efficiency,
gas-to-dust ratio, and stellar environment can all alter the observable
gas and dust signatures. The significance of the present candidate is
instead that it provides a direct route toward linking atomic
spectroscopy with the mineral-vapor process responsible for rocky
planet disintegration.}

\subsection{High-Resolution Test of the O~{\sc i} Interpretation}
\label{sec:highres}

{The most direct test of the interpretation is phase-resolved
high-resolution spectroscopy of the 7774~\AA\ triplet. At the
ALPY~600 resolution, all three O~{\sc i} components and the stellar
photospheric line are blended. Resolving the triplet would allow their
relative depths, velocity structure, and orbital-phase dependence to
be tested independently.}

{An immediate archival opportunity is provided by the seven NEID
spectra obtained by \citet{Hon2025} with the WIYN 3.5~m telescope.
NEID covers the full triplet at $R\sim120{,}000$, corresponding near
7774~\AA\ to approximately $0.065$~\AA\ or
$2.5~{\rm km\,s^{-1}}$ resolution. The individual components,
separated by 2.23 and 1.22~\AA, are therefore fully resolved.}

{The individual exposures rather than a phase-averaged stellar stack
are required for this test. A planetary signal should vary with
orbital phase, whereas photospheric absorption remains tied to the
stellar spectrum, telluric absorption remains fixed in the terrestrial
frame before barycentric correction, and airglow appears as narrow
additive emission. A convincing confirmation would be the recovery of
phase-correlated absorption in multiple components of the triplet with
a common velocity behavior inconsistent with stellar
center-to-limb variation or terrestrial contamination.}

{Conversely, sufficiently sensitive high-resolution spectra showing no
phase-dependent triplet absorption would place a strong limit on the
excited O~{\sc i} column and challenge the planetary interpretation
of the low-resolution feature. Dedicated spectroscopy obtained
continuously before, during, and after a dust-tail transit, together
with simultaneous photometry, would provide the strongest test because
the dust-production rate of BD+05~4868~Ab varies substantially from
epoch to epoch.}

\section{Conclusion}
\label{sec:conclusion}

{We have presented time-resolved, low-resolution optical spectroscopy
and complementary ground-based photometry of the disintegrating rocky
planet BD+05~4868~Ab. The observations reveal a transit-correlated
absorption feature near the unresolved O~{\sc i} 7774~\AA\ triplet.
We interpret this feature as candidate absorption from neutral oxygen
in an escaping mineral-vapor component and test that interpretation
with a one-zone NLTE forward model. Our principal conclusions are as
follows.}

\begin{enumerate}

\item
{The in-transit spectrum contains additional absorption near the
O~{\sc i} triplet at 7771.94, 7774.17, and 7775.39~\AA\ relative to
the out-of-transit baseline. At the ALPY~600 resolving power of
$R\simeq600$, the triplet is unresolved within an approximately
13~\AA\ instrumental profile. The differential feature has an
equivalent width of approximately
$W_\lambda\simeq1.4\pm0.2$~\AA, a maximum differential depth of
order 9\%, and an apparent width of approximately 13--16~\AA.}

\item
{An oxygen-bearing gaseous component is a natural consequence of the
rock-vapor outflow expected from BD+05~4868~Ab. Vaporization of
silicate and oxide material can supply O$_2$, O, SiO, Mg, Fe, and
other refractory species, linking the candidate atomic absorption to
the same escaping material that ultimately produces the observed
dust tail.}

\item
{The O~{\sc i} 7774~\AA\ triplet requires non-LTE excitation under the
conditions of this system. The $3s\,{}^5S^\circ_2$ lower level lies
9.146~eV above the ground state and cannot be appreciably populated
thermally at the reported planetary temperature. Photoionization
followed by O$^+$ recombination and radiative cascade provides a
natural pathway into the excited quintet system.}

\item {The NLTE forward model quantitatively reproduces the observed
absorption. For a representative model adopting
$h_{\rm gas}=0.25R_\star$, $T_e=7500$~K, and
$L_{\rm EUV}/L_{\rm bol}=3\times10^{-6}$, the projected geometry
gives $C_f\simeq0.315$, and reproducing
$W_\lambda\simeq1.4$~\AA\ requires
$\dot M_{\rm gas}/\dot M_{\rm dust}\simeq5.1$. After convolution with the $R\simeq600$ ALPY~600 instrumental
response, the same model produces an apparent differential depth of
approximately 9.7\% and a full width at half maximum (FWHM) of
approximately 13.5~\AA, in close agreement with the observed scale of
the unresolved feature. This width is comparable to the approximately
13~\AA\ instrumental resolution near 7774~\AA, so the modeled feature
would remain unresolved in the present spectra. The agreement in
equivalent width, depth, and instrumentally broadened width demonstrates
that an oxygen-bearing NLTE outflow can quantitatively reproduce the
observed absorption.}

\item
{The wavelength coincidence, transit dependence, rock-vapor chemistry,
and quantitative NLTE agreement together make escaping
O~{\sc i} our preferred interpretation of the 7774~\AA\ feature.
Because the present spectra do not resolve the triplet from the
intrinsic stellar O~{\sc i} line, we retain the designation
``candidate O~{\sc i}.'' Stellar center-to-limb and line-profile
variations remain the most important alternative. High-resolution,
phase-resolved spectroscopy capable of resolving multiple triplet
components provides a direct test of the interpretation.}

\end{enumerate}

{The existing $R\sim120{,}000$ NEID spectra of BD+05~4868~A provide
a particularly valuable opportunity for such a test. Confirmation of
phase-dependent absorption in multiple O~{\sc i} triplet components
would establish an atomic-gas counterpart to the dusty effluent and
open a complementary route to studying the composition and physical
evolution of disintegrating rocky planets. BD+05~4868~Ab is therefore
a promising system for connecting the mineral vapor, atomic gas, and
condensed dust phases of catastrophic rocky-planet mass loss.}

\begin{acknowledgments}
Spectroscopic data reduction made use of ISIS and the ALPY~600
spectral reference library distributed for use with that software.
Photometric calibration and differential aperture photometry were
performed with AstroImageJ. Python-based numerical calculations were
used for spectral analysis, reconstruction of the published dust-tail
model, development of the one-zone NLTE O~{\sc i} forward model,
instrumental convolution, parameter-space calculations, and figure
generation. These calculations made use of NumPy, SciPy, pandas, and
Matplotlib.

Atomic transition wavelengths and level energies were obtained in part
from the NIST Atomic Spectra Database. This research also made use of
the NASA Exoplanet Archive, which is operated by the California
Institute of Technology under contract with the National Aeronautics
and Space Administration under the Exoplanet Exploration Program.

We acknowledge support from the Department of Earth and Environmental
Systems at Indiana State University, the Door Peninsula Astronomical
Society, the Summer Undergraduate Research Experiences (SURE)
program, Dr.\ Jen Latimer, and Indiana State University's State
Advantage Grant Program.
\end{acknowledgments}

\software{
ISIS \citep{Buil2020},
AstroImageJ \citep{Collins2017},
Python,
NumPy \citep{Harris2020},
SciPy \citep{Virtanen2020},
pandas \citep{McKinney2010},
Matplotlib \citep{Hunter2007}
}

\vfill
\clearpage

\bibliographystyle{aasjournalv7}  

\bibliography{references}         

\end{document}